\documentclass[twocolumn,preprintnumbers,amsmath,amssymb]{revtex4}

\usepackage{graphicx}
\usepackage{dcolumn}
\usepackage{amsthm}
\usepackage{bm}

\begin{document}


\title{Unconventional Metallicity and Giant Thermopower in a \\ Strongly Interacting Two Dimensional Electron System}

\author{Vijay Narayan$^{*1,2}$, Srijit Goswami$^2$, Michael Pepper$^3$, Jonathan Griffiths$^1$, Harvey Beere$^1$, Francois Sfigakis$^1$, Geb Jones$^1$, Dave Ritchie$^1$, Arindam Ghosh$^2$}%
\affiliation{%
$^1$Cavendish Laboratory, University of Cambridge, J. J. Thomson Avenue, Cambridge CB3 0HE, United Kingdom \\ $^2$Department of Physics, Indian Institute of Science, Bangalore 560012, India \\ $^3$	Department of Electronic and Electrical Engineering, University College London, Torrington Place, London WC1E 7JE, United Kingdom}%

\date{\today}
\begin{abstract}
We present thermal and electrical transport measurements of low-density (10$^{14}$~m$^{-2}$), mesoscopic two-dimensional electron systems (2DESs) in GaAs/AlGaAs heterostructures at sub-Kelvin temperatures. We find that even in the supposedly strongly localized regime, where the electrical resistivity of the system is two orders of magnitude greater than the quantum of resistance $h/e^2$, the thermopower decreases linearly with temperature indicating metallicity. Remarkably, the magnitude of the thermopower exceeds the predicted value in noninteracting metallic 2DESs at similar carrier densities by over two orders of magnitude. Our results indicate a new quantum state and possibly a novel class of itinerant quasiparticles in dilute 2DESs at low temperatures where the Coulomb interaction plays a pivotal role.
\end{abstract}

\maketitle

\section{INTRODUCTION}

The thermopower or Seebeck coefficient $S$ of a system is the electric voltage $V_{th}$ generated in response to an imposed temperature difference $\Delta T$ across its ends. In an electron system thermally connected to a phonon bath, the primary contributions to $V_{th}$ are from ``phonon drag,'' which is driven by electron-phonon scattering, and the diffusive kinetics of the electrons themselves in order to maintain local thermal equilibrium. The latter is dominant at sufficiently low temperatures when most phonons freeze out and is called the diffusion thermopower ($S_d$). Within the semiclassical Boltzmann framework, the diffusion thermopower is given by the Mott relation~\cite{Mott} and connects it to the conductivity $\sigma$ of the system as

\begin{equation}
\label{Mottformula}
S_d \equiv \frac{V_{th}}{\Delta T} = \frac{\pi^2 k_B^2 T}{3q}\left(\frac{d\ln \sigma}{dE}\right)_{E = \mu}
\end{equation}

\noindent where $k_B$ is the Boltzmann constant, $T$ is the average electron temperature, $q$ is the charge of the carriers, $E$ is the energy, and $\mu$ is the chemical potential of the system. Thus, $S$ is sensitive to the energy-derivative of $\sigma$ and consequently various system parameters such as the electronic density of states (DOS) and the momentum relaxation time $\tau$. This sets it apart from the resistivity as a spectroscopic tool and the physics it probes. For instance, the $T$~dependence of $S$ is markedly different for metals, Anderson insulators, and gapped insulators: For metals, where there is a continuous DOS and free charge carriers, $S(T)$ decreases to zero linearly as $T \rightarrow 0$; for 2D Anderson insulators which are characterized by a sharp mobility edge separating the conducting (extended) and nonconducting (localized) states in an otherwise continuous DOS, $S(T)$ varies as $T^{1/3}$; in the Efros-Shklovskii regime, where Coulomb interactions cause the opening of a soft gap at the Fermi energy in the DOS, $S(T) \rightarrow$ constant as $ T \rightarrow 0$; and finally, for hard gapped insulators $S(T)$ diverges as $1/T$. These distinctions are far less pronounced in the resistivity $\rho(T)$. Another virtue of $S$ is its equivalence to the entropy per carrier, thus providing a probe of this abstract quantity. We note here that the Mott formula [Eq.~(\ref{Mottformula})] was originally derived for a strongly degenerate, noninteracting metal and hence is not expected to be valid when strong interactions are present.

\begin{figure}
	\centering
		\includegraphics[width=3.25in]{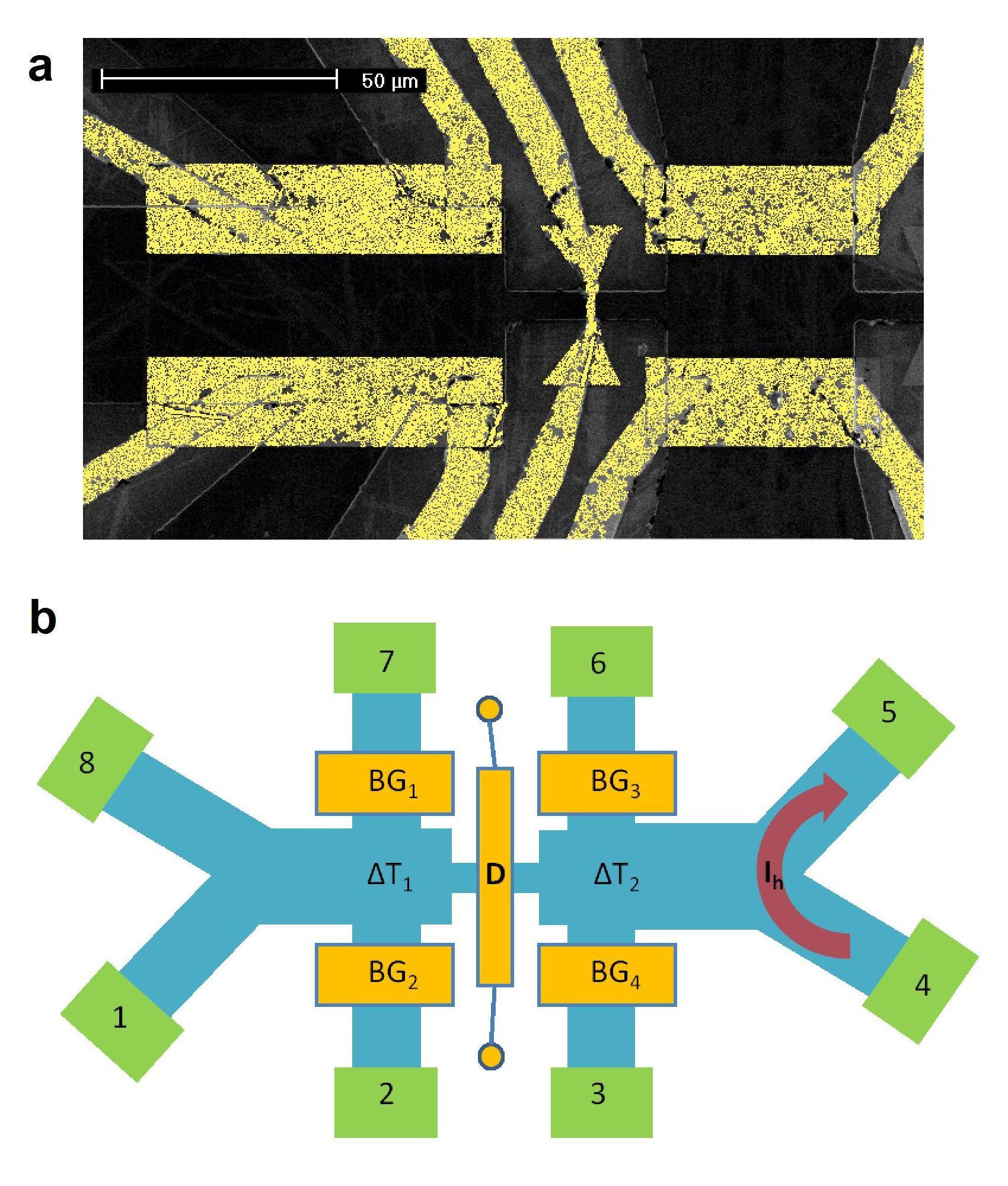}
	 \caption{(Color online) (a) False-color SEM image of a typical device. (b) Schematic of the device layout. D denotes the mesoscopic device, BG$_{1-4}$ are macroscopic bar gates used for thermometry (see Sec.~\ref{BGThermometry}), and 1--8 represent ohmic contacts. A heating current $I_h$ establishes a temperature gradient along the length of the device.}
	\label{Device}
\end{figure}

The theromopower has proved to be a powerful tool to probe many-body phases in strongly interacting 2DESs, particularly those involving transitions from insulating to either metallic or fractional quantum Hall liquid phases. In the presence of strong transverse magnetic fields, the thermopower has been employed to investigate the nature of the insulating state near the $\nu = 3/2$ fractional quantum Hall state~\cite{PossanziniPRL2003}, the existence of an energy gap at the $\nu = 5/2$ fractional quantum Hall state~\cite{Chickering2010}, and the possibility of a Wigner crystalline ground state near $\nu = 1$ in a bilayer GaAs/AlGaAs hole system~\cite{FanielPRL2005}. The possibility of a metallic ground state in low-density 2DESs at zero magnetic field, where the metallicity is driven purely by Coulomb interactions, has however been far more controversial~\cite{AbrahamsRMP2001}. Several attempts to identify such a metal have been made with thermopower as a probe in the apparently metallic regime, notably in low-density 2D electron or hole systems~\cite{MoldovanPRL2000, SrijitPRL} and high-mobility Si MOSFETs~\cite{FletcherSST2001}. While these studies indicate a definite change in the conduction mechanism and/or critical behavior in thermopower near the transition point between insulator and apparent metal, an unambiguous demonstration of a metallic state has never been achieved. Much of the uncertainty could be due to the very nature of the transition which has been suggested to be an inhomogeneity-driven, classical percolation transition~\cite{MeirPRL1999, SDSarmaPRL2005} rather than an interaction-induced one.

Recently, the possibility of a metallic phase in strongly interacting 2DESs in zero magnetic field has resurfaced through transport experiments with micron-scale 2DESs in GaAs/AlGaAs heterostructures~\cite{ArindamJPhysC2004, ArindamPRB2004, Matthias, Koushik}. The motivation behind using micron-sized or ``mesoscopic'' samples was to circumvent the influence of the long-ranged disorder known to exist in molecular-beam-epitaxy-grown GaAs/AlGaAs heterostructures~\cite{Finkelstein, Chakraborty}. This greatly reduces the tendency of the 2DES to fragment into puddles and, consequently, the chances that the transport be governed by percolation through these puddles. Indeed, at low $n_s$, Baenninger \textit{et al.}~\cite{Matthias} observed striking behavior in mesoscopic 2DESs: The $T$~dependence of the resistance weakened drastically either saturating to a finite value or decreasing with decreasing $T$ even though the absolute value of resistivity exceeded $h/e^2$ by several orders of magnitude. The aim of the present work is to investigate the nature of the DOS at the Fermi energy and many-body phenomena in these systems using thermopower measurements.

\section{EXPERIMENT}

\begin{figure}[b]
	\centering
		\includegraphics[width=3.25in]{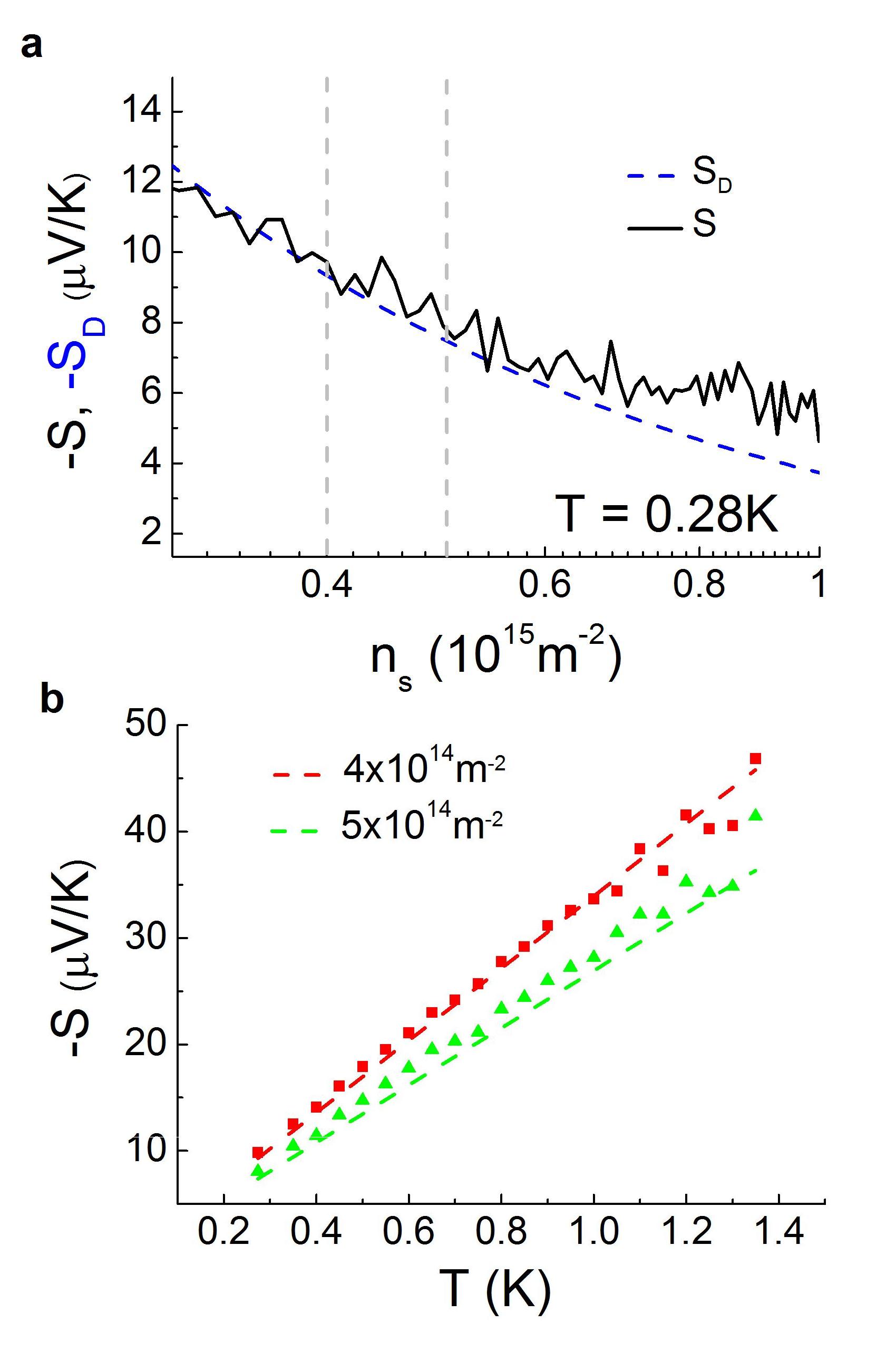}
		\caption{(Color online) Density and temperature dependence, respectively, of $S$ in the high-$n_s$ regime. The dashed lines show $S_d$ [Eq.~(\ref{Drude-Mottformula})]}
	\label{Control}
\end{figure}

Figure~\ref{Device}(a) shows a micrograph of a typical device. The devices are fabricated in Si-doped GaAs/AlGaAs heterostructures in which the distance between the dopant layer and 2DES is 40~nm. The as-grown mobility was 220~m$^2$/Vs at carrier densities of 2.2$\times$~10$^{15}$~m$^{-2}$. The mesoscopic 2DES is defined using a top-gate of dimensions $L\times W = 2\mu$m $\times$ 8$\mu$m, which enables us to tune its density $n_s$ by applying a gate voltage $V_g$. The $n_s$-$V_g$ calibration is obtained via an edge state reflection technique~\cite{MatthiasPRB2005}. We measure the 2DES resistivity $\rho_{2DES}$ in a 4-probe setup by passing a small excitation current $I_{ex}$ = 100~pA at $f$ = 7~Hz and detecting the output $V_{ex}$ using a lock-in amplifier. To measure $S$ we impose a temperature difference $\Delta T$ across the device by means of a heating current $I_h$ = 4-5~$\mu$A at $f_h$ = 11~Hz and detect the thermovoltage $V_{th}$ at 2$f_h$. $\Delta T$ across the device is measured using large ($\sim$ 10$\mu$m) bar gates following references~\cite{Appleyard} and~\cite{Chickering2009} (see Sec.~\ref{BGThermometry} for further details). In the results reported here $\Delta T$ never exceeds 20~mK (see Fig.~\ref{DeltaT}). In our experiments $\rho_{2DES}$, $V_{th}$ and $\Delta T$ are all measured simultaneously. We perform measurements on three devices and observe similar results in all. 

\subsection{Thermometry using bar gates}
\label{BGThermometry}

We measure the local electron temperature by measuring the thermovoltage between large 2DESs ($\approx$ 10~$\mu$m $\times$ 30~$\mu$m) defined by metallic gates. The layout of the bar gates can be seen in Figs.~\ref{Device}(a) and \ref{Device}(b). At high $n_s$ when interaction or localization effects are negligible and the 2DES is well described as a noninteracting, Drude-like metal, Eq.~(\ref{Mottformula}) reduces to,

\begin{equation}
\label{Drude-Mottformula}
S_d = -\frac{\pi k_B^2Tm}{3e\hbar^2}\frac{1 + \alpha}{n}
\end{equation}

Here $k_B$ is Boltzmann's constant, $T$ is the average temperature $\equiv (T_e + T_L)/2$ with $T_e$ and $T_L$ the electron and lattice temperatures, respectively, $m$ is the effective electron mass in GaAs , $-e$ is the electronic charge, $\hbar$ is Planck's constant divided by $2\pi$, $n$ is the 2DEG density, and $\alpha \equiv \frac{n}{\tau}\frac{d\tau}{dn}$, where $\tau$ is the momentum relaxation time. To measure $\Delta T_1$, say [see Fig.~\ref{Device}b], where the local electron temperature is $T_{e1}$, we differentially bias BG$_1$ and BG$_2$ and detect the the signal $V_{\Delta T_1}$ between contacts 2 and 7 at twice the heating frequency. The difference in thermopowers between BG$_1$ and BG$_2$, $\Delta S_d \equiv V_{\Delta T_1}/\Delta T = V_{\Delta T_1}/(T_{e1} - T_L)$. Substituting the expression for $S_d$ from Eq.~(\ref{Drude-Mottformula}) we obtain

\begin{equation}
\label{Te}
T_{e1}^2 = V_{\Delta T_1}\frac{3e\hbar^2}{\pi k_B^2 m (1 + \alpha)}\left(\frac{1}{n_1} - \frac{1}{n_2}\right)^{-1} + T_L^2
\end{equation}

\noindent where $n_1$ and $n_2$ are the 2DEG densities beneath BG$_1$ and BG$_2$, respectively, and are obtained via an edge-state reflection technique~\cite{MatthiasPRB2005}. $T_L$ is obtained from a Ru$_2$O thermometer attached close to the sample. A similar procedure yields the value of $T_{e2}$ from which $\Delta T$, the temperature difference across D, the mesoscopic device under study, is obtained as $T_{e2} - T_{e1}$.

\begin{figure}
	\centering
		\includegraphics[width=3.25in]{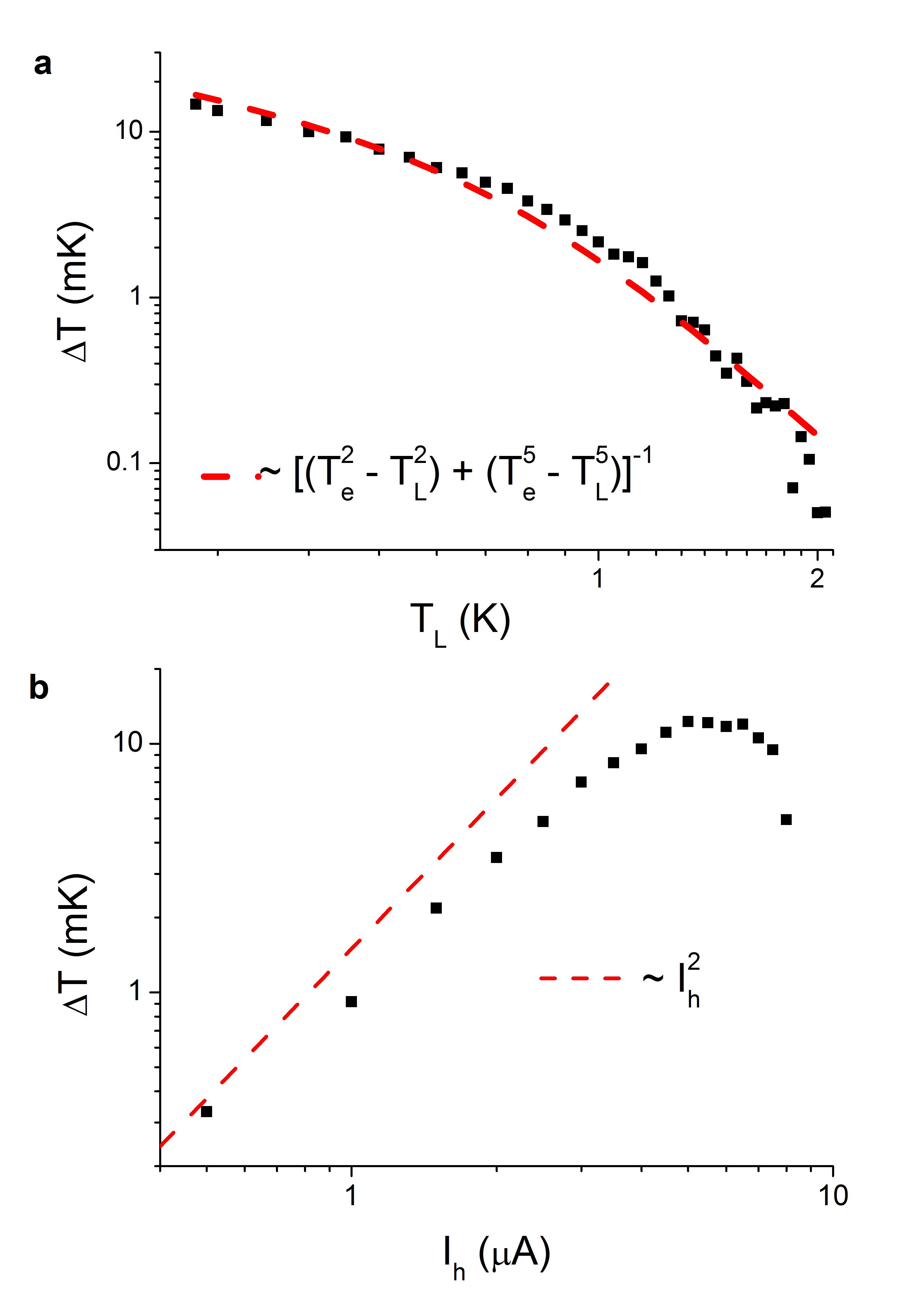}
		\caption{(Color online) (a) The figure shows the variation of $\Delta T$ with the lattice temperature $T_L$. Also shown is the qualitative dependence of $\Delta T$ expected due to heat dissipation by piezoelectric electron-phonon scattering and to the ohmic contacts. (b) The figure shows that as $I_h$ is increased $\Delta T$ increases approximately quadratically before nonlinearities appear at large $I_h$.}
	\label{DeltaT}
\end{figure}

In order to calibrate our setup we have first carried out detailed thermopower measurements when D is tuned to relatively high $n_s$ where $\rho_{2DES} < h/e^2 $ and interaction or localization effects are negligible. Here the 2DES is well described as a noninteracting, Drude-like metal and its thermopower given by Eq.~(\ref{Drude-Mottformula}). Figure~\ref{Control}(a) shows the dependence of $S$ on $n_s$ in the regime $n_s > 3\times10^{14}$~m$^{-2}$, measured at base temperature, $T = 0.28$~K. The dashed line represents $S_d$ for $\alpha = 1$. In Fig.~\ref{Control}(b), we present the $T$ dependence of $S$ at two values of $n_s$ [$= 4$ and $5 \times 10^{14}$~m$^{-2}$, denoted by the vertical dashed lines in Fig.~\ref{Control}(a)]. At both $n_s$ we find $S$ to be well described by Eq.~(\ref{Drude-Mottformula}). Apart from a consistency check for the measurement process, Figs.~\ref{Control}(a) and \ref{Control}(b) also confirm that $S$ is dominated by the diffusive component over this range of $T$ ($\lesssim 1.5$~K), and that the phonon drag contribution to $S$ is negligible. This is not surprising as the hot-electron technique greatly reduces the phonon drag contribution as was demonstrated by Chickering {\it et al.} in Ref.~\cite{Chickering2009}. 

A second validation of our measurement technique is shown in Fig.~\ref{DeltaT}(a). We see that $\Delta T$ shows two distinct dependencies on $T$: A slower decay when $T \leq 0.6K$ and a faster decay when $T > 0.6K$. The behavior of $\Delta T$ can be accounted for simply by a combination of power dissipated to the ohmic contacts, $P_O$, and through piezoelectric electron-phonon coupling, $P_{ep}$. The net power dissipated $P \equiv P_O + P_{ep} = I_h^2R$. Here $I_h$ is the (constant) heating current and $R$ the resistance of the heating channel [shown as red arrow in Fig.~\ref{Device}(b)]. In the range of temperatures explored, $R$ is completely dominated by impurity scattering and hence shows no temperature dependence. Consequently, $P$ is independent of $T$. Assuming $P_O =  K_1(T_e^2 - T_L^2)$ and $P_{ep} = K_2(T_e^5 - T_L^5)$~\cite{MaPRB1991} where $K_1$ and $K_2$ are system-dependent constants, we obtain

\begin{widetext}
\begin{equation}
\label{P}
\Delta T \equiv T_e - T_L \propto (K_1(T_e + T_L) + K_2(T_e^4 + T_e^3T_L + T_e^2T_L^2 + T_eT_L^3 + T_L^4))^{-1}
\end{equation}
\end{widetext}

The qualitative behavior described by Eq.~(\ref{P}) is shown as a broken line in Fig.~\ref{DeltaT}(a) and is seen to adequately describe the observed dependence. We mention here that even in the absence of a voltage on the gate finger D in Fig.~\ref{Device}(b), $\Delta T_1$ = 0; i.e., the electrons relax to $T_L$ over the distance between the heating channel and cold end of the device. Consequently, $\Delta T \equiv T_{e2} - T_L$ does not change as the 2DES is tuned between the high- and low-density regimes. A third check is shown in Fig.~\ref{DeltaT}(b) where we plot $\Delta T \equiv T_e - T_L$ against the heating current $I_h$. We see that as expected, $\Delta T$ grows as $\approx I_h^2$ before saturating, presumably due to nonlinearities arising from the large values of $I_h$. 

\section{RESULTS}

\begin{figure}[b]
	\centering
		\includegraphics[width=3.25in]{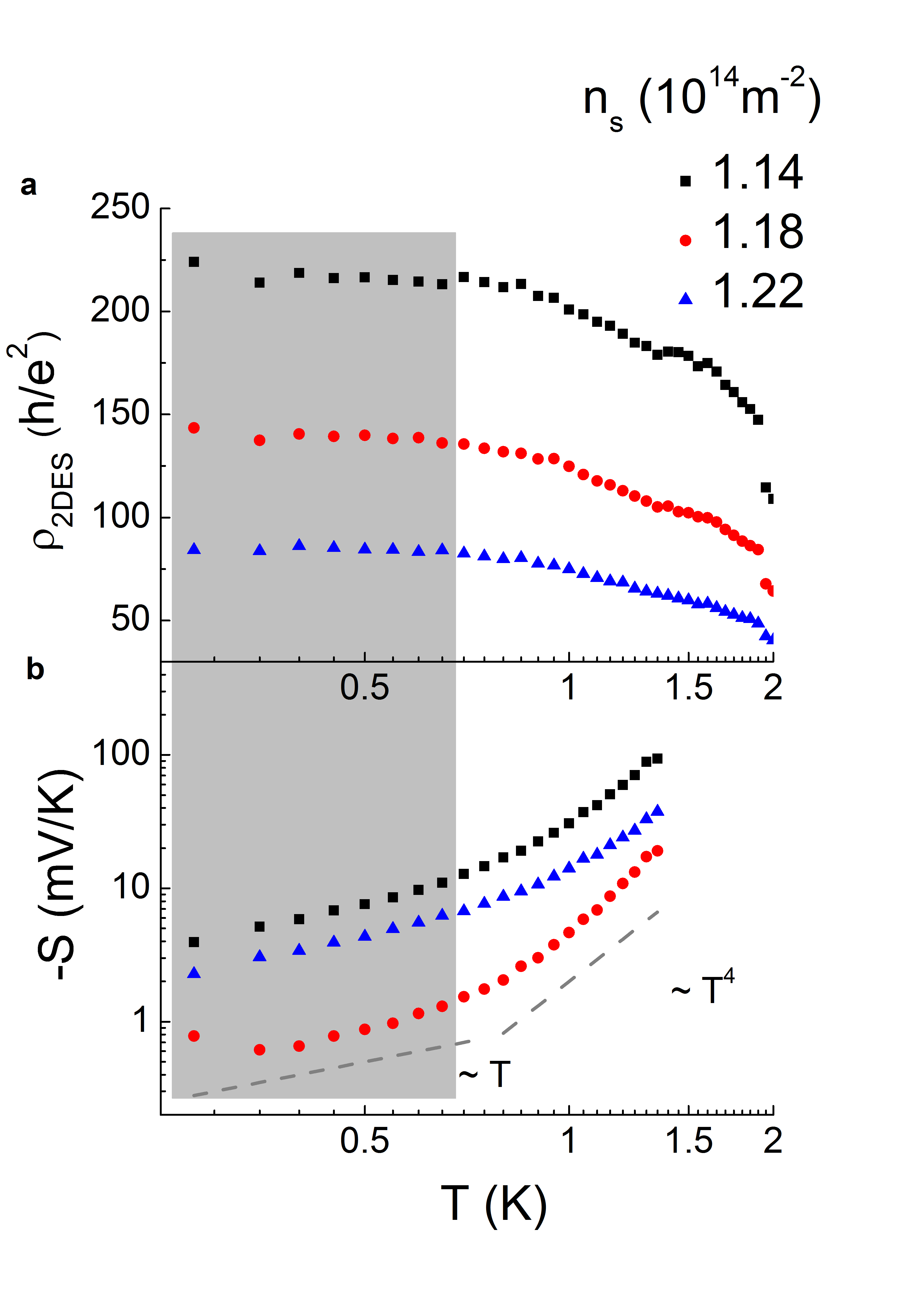}
	 \caption{(Color online) (a) In the highly resistive regime $\rho_{2DES}$ becomes almost $T$ independent below $\sim 0.8$~K. (b) $S$ increases linearly below  $T \lesssim$ 0.8~K beyond which it increases as $T^4$, in agreement with earlier reports of thermopower in GaAs/AlGaAs heterostructures~(Ref.~\cite{FanielPRL2005}).}
	\label{TempDependence}
\end{figure}

\begin{figure*}
	\centering
		\includegraphics[width=6.5in]{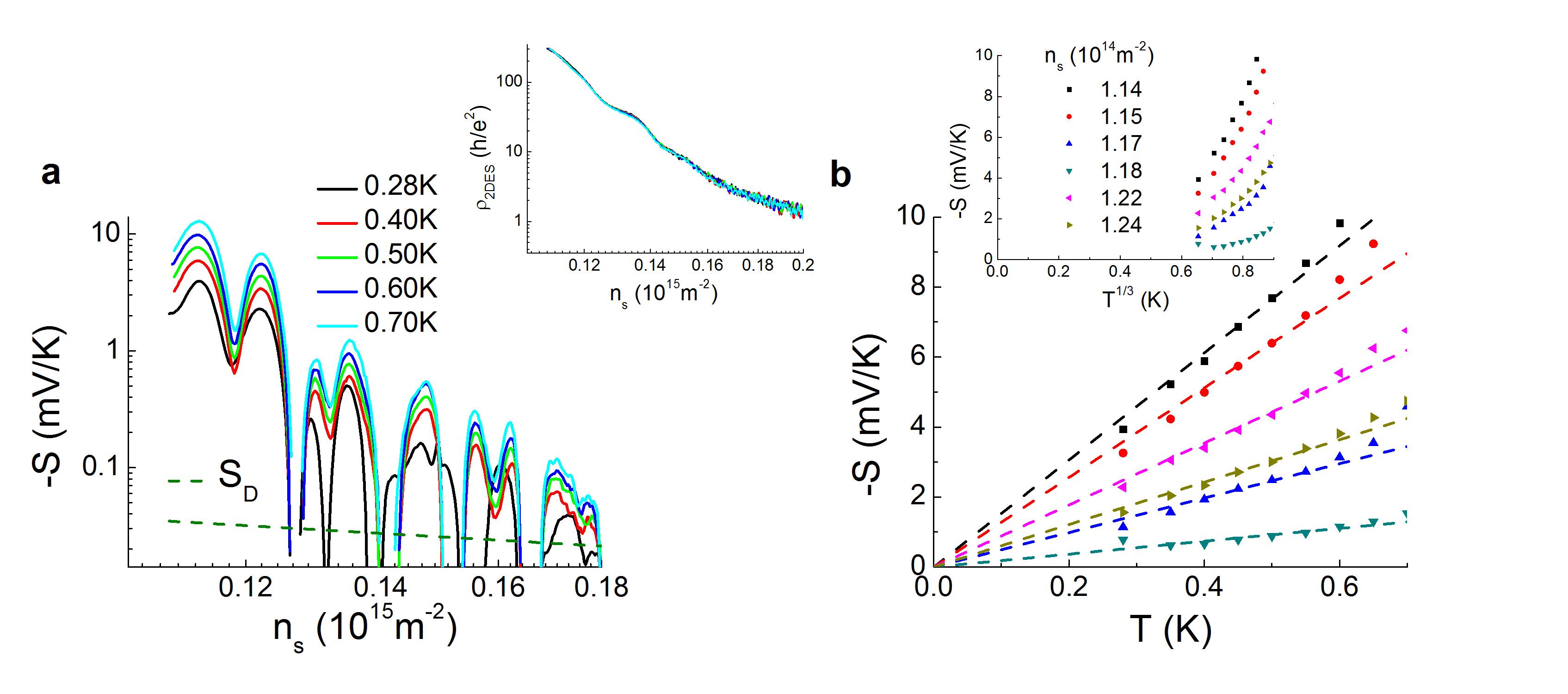}
	 \caption{(Color online) (a) $S$ vs $n_s$ for 0.28~K $< T <$ 0.7~K. The broken green line shows $S_d$ [Eq.~(\ref{Drude-Mottformula})] at 0.28~K. Inset: $\rho_{2DES}$ vs $n_s$ at the same $T$ values; there is little $T$ dependence in this range. (b) Low-$T$ linear variation of $S$. Inset: Descriptions based on variable-ranged hopping, where $S$ is expected to decay to zero as $T^{1/3}$, do not adequately describe the observed data.}
	\label{DensityDependence}
\end{figure*}

As $n_s$ in the mesoscopic region is reduced by making the top-gate voltage more negative, $\rho_{2DES}$ increases rapidly below $n_s \sim 2 - 3\times10^{14}$~m$^{-2}$ (inset Fig.~\ref{DensityDependence} and Refs.~\cite{Matthias} and~\cite{Koushik}). The $T$ and $n_s$ dependencies of $\rho_{2DES}$ in such low-$n_s$, mesoscopic 2DESs are very different from those of their macroscopic counterparts~\cite{Matthias, Koushik}. In Fig.~\ref{TempDependence}(a) we see, similar to what was observed in Ref.~\cite{Matthias}, that at low $n_s$ values, where the 2DES is expected to be localized, $\rho_{2DES}(T)$ does not diverge as $T \rightarrow$ 0 but saturates at values $\gg h/e^2$ below $\approx$ 0.8~K, indicating noninsulating behavior. The precise mechanism for resistivity saturation is debated, ranging from quantum tunneling between multiple electron puddles~\cite{TripathiKennett, NeilsonHamilton} to defect migration in a spontaneously broken symmetry phase~\cite{Matthias, Koushik}. Figure~\ref{TempDependence}(b) shows $S$ as a function of $T$ at the same $n_s$ as Fig.~\ref{TempDependence}(a). Its behavior is, however, generic to all $n_s$ in the ``localized'' regime: At $n_s \lesssim 2\times10^{14}$~m$^{-2}$, we find that $S$ always decreases with decreasing $T$ despite the fact that $\rho_{2DES} \gg h/e^2$. The nature of its decrease depends on the range: At $T \lesssim 0.8$~K, $S$ varies linearly with $T$, as in the case of a metal. This linear dependence is observed down to the lowest measured $n_s$ where the electrical resistivity can be as high as $200 - 300\times h/e^2$. When $T > 0.8$~K, $S$ is seen to grow at a much faster rate, $S \sim T^4$, as a result of which $S$ reaches extremely large values $\sim$100~mV/K at 1.4~K. The $T^4$ dependence of $S$ in GaAs/AlGaAs heterojunctions at low $T$ is usually understood in terms of phonon drag~\cite{LyoPRB1988, FanielPRL2005}, but its absence at the same $T$ at high $n_s$ in our devices (see Fig.~\ref{Control}) suggests that the present instance warrants closer inspection. We note, however, that screening by free carriers in GaAs/AlGaAs heterostructures can significantly alter the polaron (electron-LO phonon) binding energy and scattering rate~\cite{SDSarmaPRB1985}, and it is possible that the absence of free electrons in the low-$n_s$ regime allows phonon drag to set in at lower $T$.

Figure~\ref{DensityDependence}(a) shows the $n_s$ dependence of $S$ at $T = 0.28$~K. There are two salient features which need to be noted: First, while $S$ increases with decreasing $n_s$, the decrease is much stronger than the $\sim 1/n_s$ behavior expected from Eq.~(\ref{Drude-Mottformula}); $S$ increases to about two orders of magnitude above the Mott value at the lowest $n_s$. Second, the increase in $S$ is oscillatory rather than smooth, even changing sign occasionally, although the detailed nature of oscillations is highly device dependent. Coulomb blockade due to electron puddles in an inhomogeneous charge distribution could lead to oscillatory thermopower~\cite{ScheibnerPRL2005}, but such a scenario is unlikely because (1) in spite of occasional change in sign, the oscillations are primarily one sided (negative) and very asymmetric around zero, and (2) none of this structure is seen in $\rho_{2DES}$ [shown in inset to Fig.~\ref{DensityDependence}(a)] which grows monotonically and is, by and large, featureless. In fact, the energy derivative of $\rho_{2DES}(n_s)$ fails to account for the oscillations (see Sec.~\ref{MottBreakdown}) both qualitatively and in magnitude, which indicates a breakdown of the semiclassical Mott picture~[Eq.~(\ref{Mottformula})]. While the precise origin of the oscillations remains unexplained, a many-body interaction effect, for example a reentrant order-disorder transition in a broken-symmetry phase, cannot be ruled out~\cite{Arti}.

Figure~\ref{DensityDependence}(b) embodies the key result of this work. We inspect the linear regime of $S(T)$ and show linear fits to the data that pass through the origin. This suggests the absence of any gap, hard or Efros-Shklovskii type, at the Fermi energy. In the inset we see that the same data plotted against $T^{1/3}$ cannot be described as a line going through the origin, showing clearly that the observed behavior is not consistent with Mott-type variable-ranged hopping in an Anderson insulator. We emphasize the difference between our result and earlier thermopower results near the apparent metal-insulator transition in macroscopic 2D systems: Apart from this being the first clear observation of metal-like thermopower in the low-$n_s$ regime, we find this behavior to exist over an extended range of $n_s$ and $\rho_{2DES}$ (up to $\sim 300\times h/e^2$ which is the upper limit of our measurements), unlike previous studies which explore only the critical regime ($\rho_{2DES} \sim 0.1 - 1\times h/e^2$).

\begin{figure}[b]
	\centering
		\includegraphics[width=3.25in]{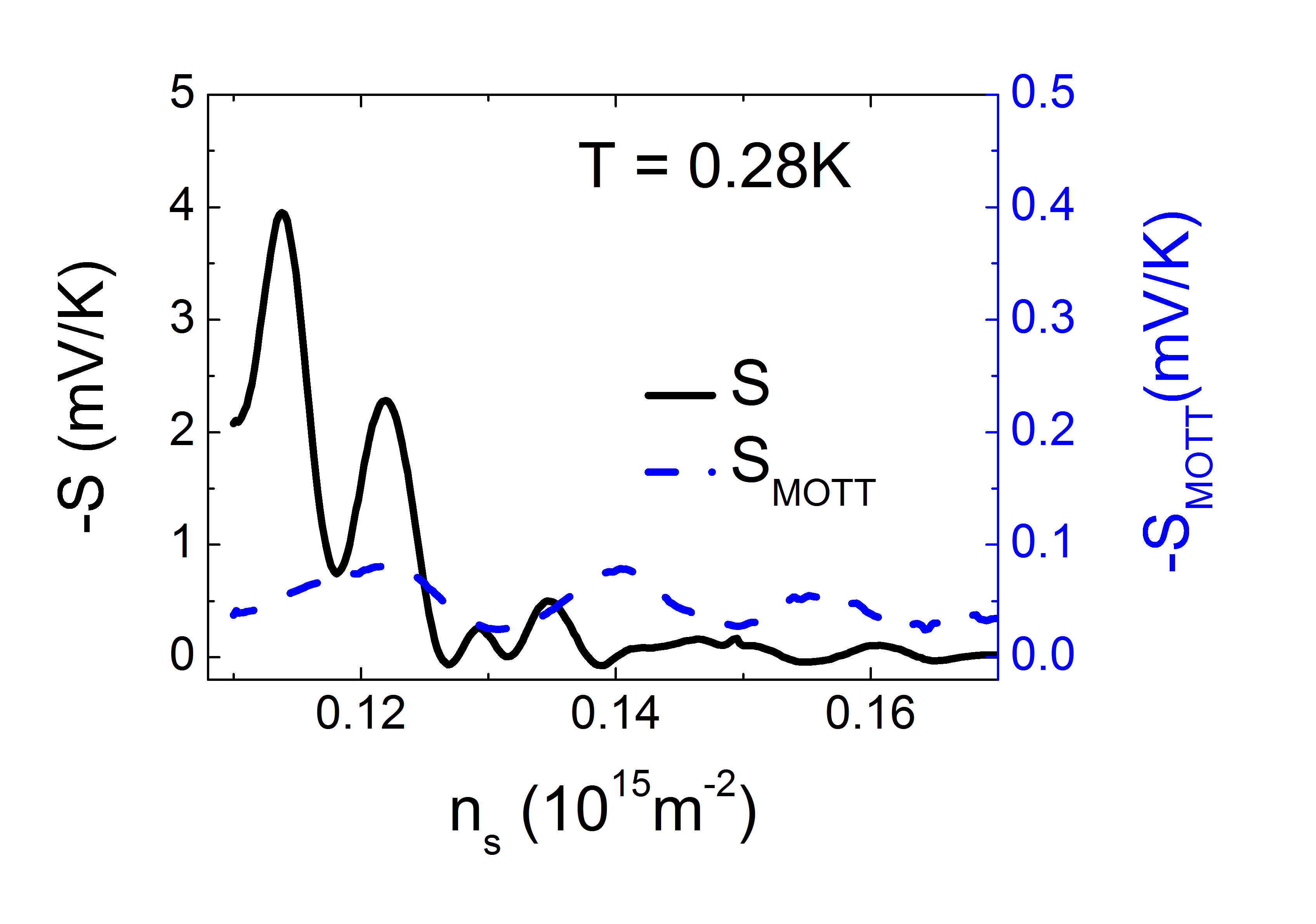}
		\caption{(Color online) The figure shows $S$ and $S_{MOTT}$~[Eq.~(\ref{Mottformula})] at $T = 0.28$~K. The $S$ trace is also shown in Fig.~\ref{DensityDependence}(a). Note that the left and right vertical axes differ by a factor of 10.}
	\label{Mottdisagreement}
\end{figure}

\subsection{Breakdown of the Mott relation}
\label{MottBreakdown}

In this section we show that the electrical and thermal transport in the low-$n_s$, mesoscopic 2DES are not simply related by the Mott formula [Eq.~(\ref{Mottformula})]. To make a quantitative comparison with the Mott result applied to our system, we evaluate the energy-derivative of $\ln\sigma$ using the Hartree expression for the energy of the 2DES

\begin{equation}
\label{2DESenergy}
E \equiv E_K + E_C = \frac{\hbar^2 \pi n_s}{m} + \frac{\hbar^2 \sqrt{\pi n_s}}{ma_B}
\end{equation}

\noindent where $a_B$ is the effective Bohr radius in GaAs $\approx 11$~nm. Using Eq.~(\ref{2DESenergy}) the expression for $S_{MOTT}$ reduces to:

\begin{equation}
\label{SMOTT}
S_{MOTT} = \frac{\pi k_B^2 T m}{3e\hbar^2}(1 + r_s/2)^{-1}\frac{d\ln \rho}{dn}
\end{equation}

In Fig.~\ref{Mottdisagreement} we compare the measured thermopower $S$ at $T = 0.28$~K to that expected from the Mott formula $S_{MOTT}$ [the former is also presented in Fig.~\ref{DensityDependence}(a)]. We note that $S$ and $S_{MOTT}$ disagree qualitatively and quantitatively. First, $S$ is nearly two orders of magnitude larger than $S_{MOTT}$ and second, though $S_{MOTT}$ captures some of the broad features in $S$, there is additional structure in the latter. The same is true for both of the other devices measured.

\begin{figure}
	\centering
		\includegraphics[width=3.25in]{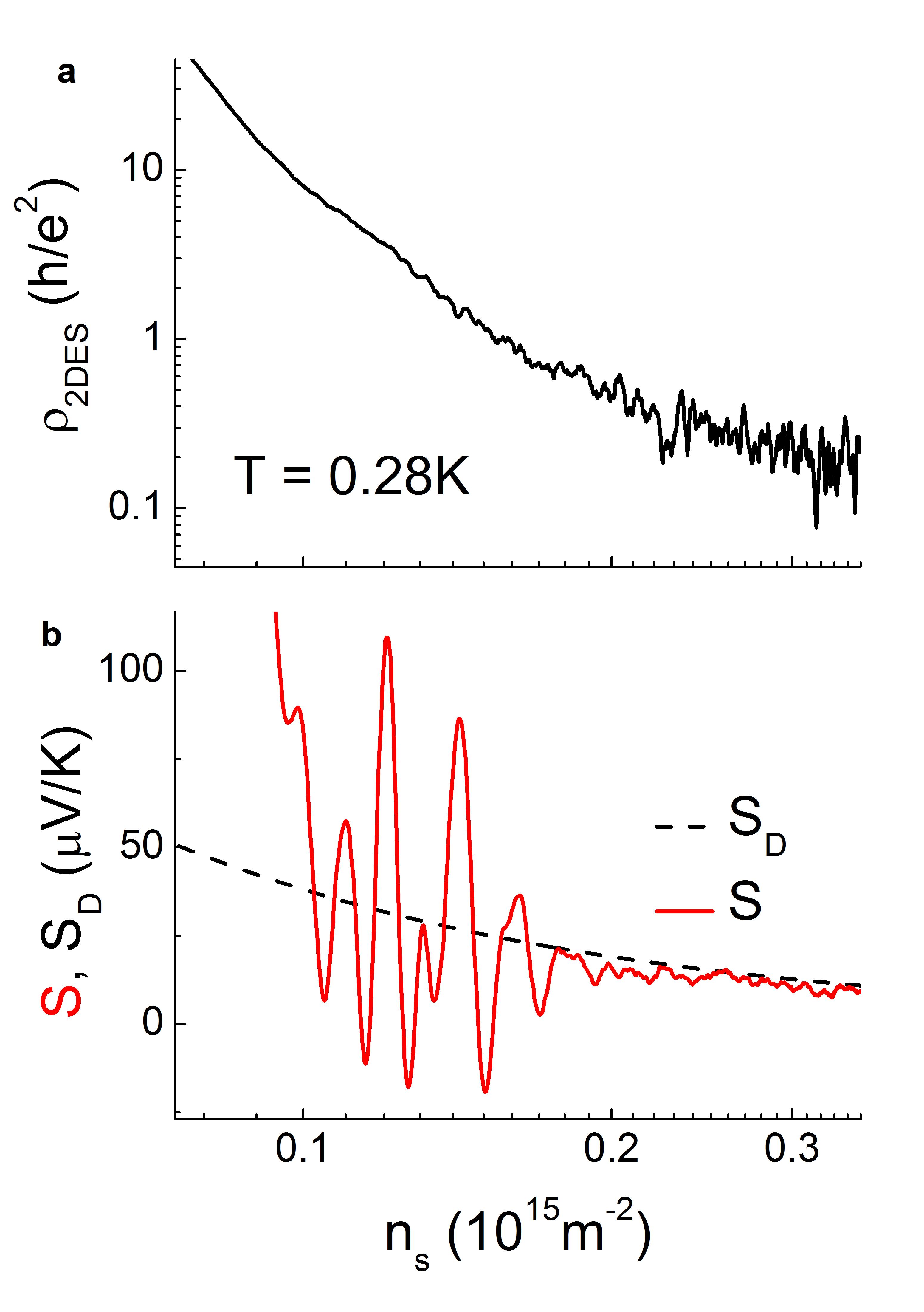}
		\caption{(Color online) The figure shows $S$ and $\rho_{2DES}$ over the same $n_s$ range when $\rho_{2DES} \sim h/e^2$. The oscillations in $S$ are strikingly present even at these low resistivities.}
	\label{high_Ns}
\end{figure}

In Fig.~\ref{high_Ns} we see that the oscillations in $S$ are observed even when $\rho_{2DES} \approx h/e^2$. The strong oscillations and even sign changes in $S$ are unaccompanied by any corresponding structure in $\rho_{2DES}$. This strongly suggests that these oscillations are not a result of Coulomb blockade in electron puddles.

\section{DISCUSSION}

\begin{figure*}
	\centering
		\includegraphics[width=6.5in]{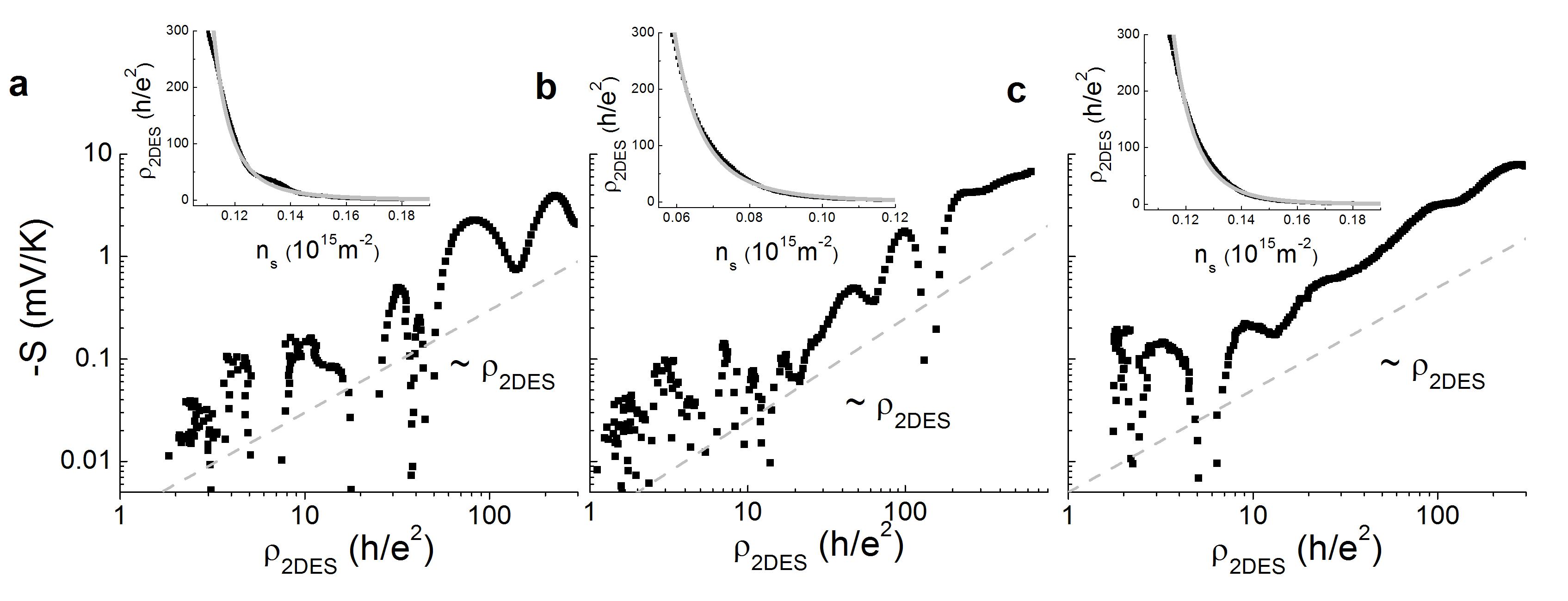}
	 \caption{(Color online) (a)--(c) In all three devices measured there is a clear linear envelope between $S(n_s)$ and $\rho_{2DES}(n_s)$ as is expected in simple Drude-like metallic system. However, the $n_s$ dependence of neither obeys a simple Drude description. The insets show that $\rho_{2DES}$ in each instance is well described by a Berezinskii-Kosterlitz-Thouless model (gray lines) suggesting that topological defects of an underlying ordered 2DES could be the mediators of transport in the dilute, mesoscopic 2DES.}
	\label{KT}
\end{figure*}

Two outstanding questions remain: First, can the nature of the metallic state be understood within an effective semiclassical Boltzmann framework? And second, why is $S$ two hundred times larger than that expected from the non-interacting model at low $n_s$? In the Drude-like metal described by Eq.~(\ref{Drude-Mottformula}), the resistivity and thermopower are both inversely proportional to the density of delocalized quasiparticles which carry heat and electricity, i.e., $S \sim 1/n_{ex} \sim \rho_{2DES}$, where $n_{ex}$ is density of delocalized quasiparticle excitations. The deviation of the observed $S$ from the Mott expectation clearly implies that $n_{ex}$ is not given simply by $n_s$. However, to see whether a Drude-like description is valid we have plotted $S$ as a function of the corresponding $\rho_{2DES}$ for three different devices at base temperature in Fig.~\ref{KT}. Remarkably, in spite of the superposed oscillatory structure, we find $S$ to be linearly proportional to $\rho_{2DES}$ over nearly three decades in $S$ in all three devices, providing strong evidence of a Drude metal-like character, but with a different nature of itinerant quasiparticles. The number of such quasiparticles can be significantly smaller than $n_s$ which can readily explain the large magnitude of both $S$ and $\rho_{2DES}$. Note that Fig.~\ref{KT} provides further evidence against percolative transport via tunneling between electron puddles that act as quantum dots since, in such cases, $S$ is expected to be proportional to the energy derivative of $\rho_{2DES}$ rather than $\rho_{2DES}$ itself~\cite{ScheibnerPRL2005}.

In the presence of strong Coulomb interactions, enhanced screening and the formation of many-body states have been suggested to lead to extended, or at least relatively less localized, electron wave functions in the ground state thereby modifying the single-electron conduction mechanism (see for example Ref.~\cite{ShepelyanskyPRL1994}). Moreover, a common effect of interactions at low temperatures is to induce broken-symmetry states such as Wigner crystals, stripe/bubble phases~\cite{KoulakovPRL1996} etc. Recently, the metal-like electrical transport observed in Refs.~\cite{Matthias} and~{Koushik} was attributed to the melting of spontaneous symmetry-broken states~\cite{Koushik}, where the melting transition proceeds through the proliferation of topological defects that form delocalized quasiparticles at sufficiently low temperatures where zero-point fluctuations dominate~\cite{AL}. The signature of the Berezinskii-Kosterlitz-Thouless (BKT) melting transition was indeed observed in all our devices [insets of Figs.~\ref{KT}(a) -- \ref{KT}(c)], where the resistivity $\rho_{2DES}$ varies as $\rho_{2DES} \propto \exp(-A/\sqrt{n_s - n_c})$, $A$ being a constant and $n_c$ the melting density~\cite{Koushik}. The corresponding delocalized excitations can be responsible for thermal transport as well, and being much smaller than the number of electrons in the system, may lead to very large thermopower. We note that though Wigner crystallisation is expected at much lower $n_s$ in disorder-free systems~\cite{TanatarCeperley}, the presence of disorder may ``pin'' the order and allow finite grain-size Wigner crystallites at much higher $n_s$~\cite{YePRL2002}.

Thus, in conclusion, the effects of the many-body Coulomb potential appear to manifest in a novel metallic phase in dilute mesoscopic 2DESs, with giant thermopower that decreases linearly as $T \rightarrow 0$. While we cannot rule out the possibility that the system turns insulating at even lower $T$, the fact that it persists down to lowest experimental $n_s$ explored here indicates that the metallic phase might be the true ground state of strongly interacting 2DESs.

\section*{ACKNOWLEDGEMENTS}

We acknowledge funding from the UK-India Education and Research Initiative (UKIERI), the Department of Science and Technology (DST), India, and the Engineering and Physical Science Research Council (EPSRC), UK. V. N. acknowledges useful discussion with Matthias Baenninger, Crispin Barnes, Nigel
Cooper, Chris Ford, and Charles Smith and Sohini Kar-Narayan, Kantimay Dasgupta, Simon Chorley, Luke Smith, Tse-Ming Chen, and Victoria Russell for help with the experiments.


\newpage

\end{document}